\documentclass[11pt]{article}

\newcommand{\blind}{0}

\makeatletter
\renewcommand\section{\@startsection {section}{1}{\z@}%
	{-3.5ex \@plus -1ex \@minus -.2ex}%
	{2.3ex \@plus.2ex}%
	{\normalfont\fontfamily{phv}\fontsize{16}{19}\bfseries}}
\renewcommand\subsection{\@startsection{subsection}{2}{\z@}%
	{-3.25ex\@plus -1ex \@minus -.2ex}%
	{1.5ex \@plus .2ex}%
	{\normalfont\fontfamily{phv}\fontsize{14}{17}\bfseries}}
\renewcommand\subsubsection{\@startsection{subsubsection}{3}{\z@}%
	{-3.25ex\@plus -1ex \@minus -.2ex}%
	{1.5ex \@plus .2ex}%
	{\normalfont\normalsize\fontfamily{phv}\fontsize{14}{17}\selectfont}}
\makeatother
\usepackage{amsmath}
\usepackage{graphicx}
\usepackage{enumerate}
\usepackage{natbib} %comment out if you do not have the package
\usepackage{url} % not crucial - just used below for the URL
\usepackage{latexsym}
\usepackage{mathptmx}
\usepackage{graphicx}
\usepackage{ulem}
\usepackage{amsfonts}
\usepackage{amssymb}
\usepackage{amsbsy}
\usepackage{color}
\usepackage{comment}
\usepackage{hyperref}
\usepackage{amsthm}
\usepackage{multirow}
\usepackage{array}
\usepackage{algorithmic}
\usepackage{algorithm}
\usepackage{booktabs}
\usepackage{endnotes}
\usepackage{mathtools}
\DeclarePairedDelimiter\ceil{\lceil}{\rceil}

\theoremstyle{plain}% Theorem-like structures
\newtheorem{theorem}{Theorem}[section]

\newtheorem{proposition}[theorem]{Proposition}
\newtheorem{assumption}[theorem]{Assumption}

\theoremstyle{definition}

\theoremstyle{remark}

\begin{document}
	
	%%%%%%%%%%%%%%%%%%%%%%%%%%%%%%%%%%%%%%%%%%%%%%%%%%%%%%%%%%%%%%%%%%%%%%%%%%%%%%
	\def\spacingset#1{\renewcommand{\baselinestretch}%
		{#1}\small\normalsize} \spacingset{1}
	%%%%%%%%%%%%%%%%%%%%%%%%%%%%%%%%%%%%%%%%%%%%%%%%%%%%%%%%%%%%%%%%%%%%%%%%%%%%%%
	
	\if0\blind
	{
		\title{\bf A Discrete Simulation Optimization Approach Towards Calibration of an Agent-based Simulation Model of Hepatitis C Virus Transmission}
		\author{Soham Das,$^a$ Navonil Mustafee,$^b$ and Varun Ramamohan$^a$ \\
			$^a$ Department of Mechanical Engineering, Indian Institute of Technology Delhi,\\
			Hauz Khas, New Delhi, 110016, India \\
			$^b$ Centre for Simulation, Analytics and Modelling, University of Exeter,\\
			Rennes Drive, Exeter, EX4 4ST, UK }
		\date{5th July, 2021}
		\maketitle
	} \fi
	
	\if1\blind
	{
		
		\title{\bf A Discrete Simulation Optimization Approach Towards Calibration of an Agent-based Simulation Model of Hepatitis C Virus Transmission}
		\author{Author information is purposely removed for double-blind review}
		
		\bigskip
		\bigskip
		\bigskip
		\begin{center}
			{\LARGE\bf \emph{IISE Transactions} \LaTeX \ Template}
		\end{center}
		\medskip
	} \fi
	\bigskip
	
	\begin{abstract}
		This study demonstrates the implementation of the stochastic ruler discrete simulation optimization method for calibrating an agent-based model (ABM) developed to simulate hepatitis C virus (HCV) transmission. The ABM simulates HCV transmission between agents interacting in multiple environments relevant for HCV transmission in the Indian context. Key outcomes of the ABM are HCV and injecting drug user (IDU) prevalences among the simulated cohort. Certain input parameters of the ABM need to be calibrated so that simulation outcomes attain values as close as possible to real-world HCV and IDU prevalences. We conceptualize the calibration process as a discrete simulation optimization problem by discretizing the calibration parameter ranges, defining an appropriate objective function, and then applying the stochastic ruler random search method to solve this problem. We also present a method that exploits the monotonic relationship between the simulation outcomes and calibration parameters to yield improved calibration solutions with lesser computational effort.
	\end{abstract}
	
	\noindent%
	{\it Keywords:} Agent-based simulation; hepatitis C virus; disease transmission model; simulation model calibration; stochastic ruler method; solution space truncation algorithm.
	
	%\newpage
	\spacingset{1.5} % DON'T change the spacing!
	
	\section{INTRODUCTION AND LITERATURE REVIEW}
	\label{sec:intro}
	
	For many complex simulation models, the data required to estimate every single input parameter of the simulation model is often not available. In this situation, one may estimate these input parameters via the process of `calibration', wherein these input parameter values are set such that the relevant simulation outputs approach as close to observed or known values of the outputs as possible \cite{law2000building}. We refer to the input parameters that are estimated as `calibration' parameters, and the known values of the simulation outcomes - which are estimated either via data collection from the field, from the published literature, or from domain experts - as `calibration targets'. In this paper, we demonstrate a discrete simulation optimization approach - specifically the use of the stochastic ruler random search method - towards calibration of a simulation model. The simulation model in question is an agent-based simulation model of the transmission of the hepatitis C virus (HCV) that we have previously developed for the Indian context  \citep{das2019agent}. %Agent-based simulation models are typically used in situations where modeling emergent behavior resulting from the interactions between components of a system is important \citep{willem2015agent}. 
	
	%Our model is an adaption of our previous HCV transmission model \citep{das2019agent} to the state of Punjab. In third world or developing countries, paucity of data often hinders efficient modelling. We find three variables whose values we do not know for the purpose of calibration using the stochastic ruler method. Then, we exploit the monotonic relationship between the prevalence outcomes and the calibration variables to develop a scheme of reducing the set of solutions based on two different fathoming conditions and then apply the original stochastic ruler method on this reduced set of solutions. Though, from the results, we cannot claim whether either method would give the same order of time reduction when applied to different simulation optimization problems or whether the effort to reduce the solution space is worthwhile. But, we were motivated that a significant improvement can motivate researchers to continue working on the two methods not only by implementing these in various experiments, but also in terms of developing the theoretical background for the methods.
	
	In recent times, the use of simulation optimization methods for simulation model calibration has increased. Genetic algorithms, evolutionary optimization, and their various adaptations have been commonly used for both discrete and continuous calibration parameter search spaces \citep{voloshin2015optimization,aghabayk2013novel}.
	Stochastic approximation methods such as simultaneous perturbation stochastic approximation have also been used for both discrete and continuous search spaces \citep{hale2015optimization}. With regard to calibration of agent-based simulation models, genetic algorithms and evolutionary optimization techniques for calibration of agent-based simulation models have been used in the studies of \cite{fabretti2013problem} and \cite{moya2021evolutionary}, respectively. \cite{fabretti2013problem} use a genetic algorithm and an adaptation of the Nelder-Mead simplex algorithm for calibration of an agent-based model for financial markets. Other than these techniques, \cite{johnson2009calibration} use a Latin hypercube design for the calibration parameter search space for the calibration of an agent-based simulation model developed to model interactions within a refugee camp. \cite{patwary2021metamodel} use a gradient-based metamodel approach to calibrate agent-based traffic simulation models. We refer readers to \cite{pietzsch2020metamodels} for a more detailed review of metamodels used in calibration of agent-based simulation models. %The study substituted a multimodal traffic network along with gradient-based metamodels which computed gradients using the derivatives of the flows in the links with respect to parameters for calibration. The metamodel was then applied for a network in Hong Kong.% For more details, we refer the reader to the study of \cite{arsenault2014comparison}, which provides a comparison between ten simulation optimization techniques applied to calibrate three simulation models of hydrological processes.
	%We found metaheuristics to be most commonly used in the literature of simulation optimization for model calibration. 
	
	We now discuss the calibration of simulation models in the healthcare space. \cite{kong2009calibration} employed genetic algorithm and simulated annealing for their calibration problem. \cite{taylor2010methods} and \cite{karnon2011calibrating} compare random search methods against the use of other simulation optimization techniques such as manual search and Nelder-Mead simplex methods, and gradient-based search. \cite{bicher2017calibration} do not use simulation optimization to calibrate their agent based simulation of rehospitalization of psychiatric patients - instead, they use the law of the iterated logarithm. Example techniques used to calibrate infectious disease transmission models include the grid search method \citep{luo2018development}, Latin hypercube modelling \citep{shrestha2017comparing}, and the genetic algorithm \citep{reiker2021machine}. \cite{hazelbag2020calibration} provides a useful reference for studies that document the use of optimization techniques to calibrate agent-based disease transmission simulation models.
	
	More generally, our search of the literature did not yield studies wherein conventional discrete simulation optimization methods such as the stochastic ruler method \cite{yan1992stochastic}, probabilistic branch and bound method \citep{norkin1998branch} and COMPASS \citep{hong2006discrete} have been used for calibration of either discrete-event or agent-based simulations - in the healthcare area or in other application areas. In this study, we demonstrate the conceptualization of the calibration process as a simulation optimization process that can be solved via the stochastic ruler random search method. %As part of this conceptualization, we discretize the calibration parameter search space and define an appropriate objective function relating the simulation outcomes, calibration targets and the calibration parameters. %Random search techniques are widely used for simulation optimization when the search space is discrete.
	
	The stochastic ruler method is one of the first discrete simulation optimization methods with provable asymptotic convergence to the global optimum, and is relatively straightforward to implement. Modifications of the stochastic ruler method have been developed to improve the convergence of the method \citep{alrefaei2005discrete}; however, we implement the original version by \cite{yan1992stochastic} as an initial proof-of-concept of this approach towards calibrating a simulation model. Further, we also demonstrate a method for using information regarding the relationship between the calibration parameters and the calibration targets to reduce the size of the search space. Applying the stochastic ruler method on this truncated calibration parameter search can yield, as we demonstrate, improved solutions while incurring a potentially lower computational cost in comparison with applying the stochastic ruler method on the original solution space. Our search of the literature did not yield another study that truncated the solution space in this manner for discrete simulation optimization problems.
	
	We now describe the agent-based simulation of HCV transmission that we use to demonstrate our model calibration approach.
	
	%The remainder of the paper is organized as follows. In Section~\ref{hcvmod}, we describe the agent-based simulation model of HCV transmission, and in Section 3, we describe the conceptualization of the calibration of this model as a discrete simulation optimization problem to be solved using the stochastic ruler method. In Section 4, we describe our search space truncation (SST) approach. We conclude in Section 5. 

	\section{THE HCV TRANSMISSION MODEL}
	\label{hcvmod}
	
	The agent-based model of HCV transmission that we develop is comprehensive, in that we incorporate all key modes of transmission of HCV in the Indian context. In India, HCV is a significant public health concern in the state of Punjab, where the prevalence of the disease is substantially higher than the average prevalence in India (3.6\% in Punjab versus 1\% in India) \citep{sood2018burden}. Because of this, and the consequent fact that data regarding HCV transmission modes (e.g., via injecting drug use) and epidemiology for Punjab is available more widely than for other states, we select Punjab as the geographical area of interest for the transmission model \citep{chakravarti2013study,ambekar2008size}.% In India, injecting drug use is the second highest contributor to the HCV load after medical procedure-based infections \citep{chakravarti2013study}. Injecting drug use is quite prevalent in Punjab and data on injecting drug practices is also available from this state \citep{ambekar2008size}. Hence, .
	
	The key modes of transmission of HCV spread in India are medical procedures (blood transfusions, surgeries, injections and dental procedures), injecting drug use, and to a lesser extent, tattooing \citep{chakravarti2013study}. In our agent-based simulation, we create representative environments for disease transmission through each of these modes. For modeling spread of HCV through medical procedures, we create a medical environment. A social interaction environment models the transmission of HCV through injecting drug use. For this, there are two processes included within the environment- conversion of non-injecting-drug-users (non-IDUs) into IDUs, and transmission of HCV due to sharing of needles between IDUs. Nearly 75\% of IDUs are in the age group of 18-29 years \citep{ambekar2008size}, and a large proportion of these are in the age group of 18 and 24 years. In Punjab, 20\% of people in the age group 18-24 years avail of higher education \citep{mha16}. Thus, to model spread among young people availing higher education, we incorporate a higher education environment. Though sexual transmission contributes very less to HCV transmissions \citep{chakravarti2013study}, we nevertheless incorporate this mode so that the simulation model can later be adapted for hepatitis B virus transmission or to study HCV/HIV comorbidities. Our ABM is the first simulation model to explicitly model the mechanisms or sub-processes of transmission of HCV through all of the above modes.
	
	The calibration targets for this simulation model are key prevalence outcomes of the simulation - HCV antibody, HCV RNA and IDU prevalence values - for which reliable estimates are available in the literature. The HCV antibody and HCV RNA observed values (which we collectively call HCV prevalence values) were taken from the study of \cite{sood2018burden}. This study documented a large cross-sectional epidemiological survey conducted in Punjab in 2014 to determine the prevalence of HCV. The IDU prevalence was estimated from the study of \cite{ambekar2008size}. The calibration process involves running the simulation for 50 years of simulation time, with daily time steps. This calibration or burn-in period was chosen as it yielded rates of increase of HCV prevalence values which were deemed suitable by our collaborating clinical expert \citep{das2019agent}.
	
	All agents in the models are placed into groups (a proxy for a `family'), where each group consists of an older pair (aged 48 years and above), young pairs (aged between 23-48 years), and children and young adults (below the age of 23 years). Note that while disease transmission through sexual interaction between pairs is included in the model, this mode is of limited interest from the calibration point of view. This is because, given the very low per-event probability of transmission of HCV \citep{osmond1993risk}, this mode of transmission has a significantly lower contribution to the prevalence of HCV \citep{chakravarti2013study,das2019agent} when compared to other transmission modes. Further, the parameter of interest - the per-event probability of transmission via this mode - is also reliably estimated.
	
	We now briefly describe the specifics of HCV transmission in each environment. %For a detailed description of the environments along with the values of HCV transmission environment parameters, we refer the reader to \cite{das2019agent}.
	
	\subsection{HCV Transmission in the Medical Environment}
	
	An agent in the simulation can visit the medical environment for a blood transfusion, a surgery, a dental procedure or to receive an injection, per the key modes of transmission documented in \cite{chakravarti2013study}. Transmission in this environment occurs as follows. 
	
	\textit{a.} Any agent in the model can visit the medical environment on a given day with probability $p_1$ calculated using equation~\ref{eqmedv}.
	\begin{equation}
		\label{eqmedv}
		p_1 = \frac{N_{inj}+N_{bt}+N_{sur}+N_{dp}}{360}
	\end{equation}
	$N_{inj}$, $N_{bt}$, $N_{sur}$ and $N_{dp}$ represent the average number of injections, average number of blood transfusions, average number of surgeries and average number of dental procedures, respectively, that an Indian person undergoes on an annual basis (that is, in 360 days).
	
	\textit{b.} At the medical environment, the number of medical professionals is estimated to be 40, based on the fact that there is approximately one doctor per 1,800 population in India (1:1800) \citep{das2019agent}. The average number of agents considered across our simulation time horizon is 75,000.
	
	\textit{c.} Based on a survey of dental clinics of India (details in \cite{das2019agent}), we assume that 50\% of medical professionals do not implement medically acceptable decontamination protocols in their workspaces. Hence, we randomly assign 20 medical professionals out of 40 as those who work in `contaminated' environments.
	
	\textit{d.} If an infected agent visits a medical professional working in a `contaminated' environment, then every uninfected agent visiting the professional after the infected agent has a probability $p_2$ of getting infected. The value of $p_2$ is found using equation~\ref{eqmedi}.
	\begin{equation}
		\label{eqmedi}
		p_2 = \frac{N_{inj} \times p_{inj} + N_{bt} \times p_{bt} + N_{sur} \times p_{sur} + N_{dp} \times p_{dp}}{N_{inj} + N_{bt} + N_{sur} + N_{dp}} 
	\end{equation}
	Here, $p_{inj}$, $p_{bt}$, $p_{sur}$ and $p_{dp}$ are the per-event probabilities of getting infected through each of the four modes of transmission within this environment.
	
	While we were able to obtain reliable estimates of $p_{inj}$, $p_{bt}$, and $p_{sur}$, we were unable to do so for $p_{dp}$. Thus, $p_2$ in effect becomes a calibration parameter, even though we can obtain a reasonable initial estimate of $p_2$ for the calibration process by assuming $p_{dp}$ to take a value between that of $p_{inj}$ and $p_{sur}$. We make this assumption because, given the nature of dental procedures, it is likely that the transmission risk is likely to be greater than that from a needle-stick injury that occurs during an injection, and likely to be lesser the transmission risk from a significantly more invasive procedure such as a surgery. This reasoning was validated by our collaborating clinical expert, as documented in \cite{das2019agent}.
	
	\subsection{HCV Transmission in the Social Interaction Environment}
	Two types of interactions occur in this environment: interactions between IDUs, and interactions between IDUs and non-IDUs. These are described below.
	
	\textit{a.} We calculate the probability of an IDU visiting the social interaction environment, $p_3$, using equation~\ref{eqiduf} below.
	\begin{equation}
		\label{eqiduf}
		p_3 = \frac{\sum\limits_{districts} N_{district} \times f_{district}}{\sum\limits_{districts} N_{district}} 
	\end{equation}
	\cite{ambekar2008size} studied IDU characteristics in certain districts of Punjab, and reported the weekly frequency of injecting drugs for IDUs in these districts. $N_{district}$ refers to the population of the districts of Punjab studied in \cite{ambekar2008size}, and $f_{district}$ refers to the weekly frequency of injecting drugs for IDUs in these districts.
	
	\textit{b.} The daily probability of a non-IDU going to the social interaction environment was assumed to be $\frac{1}{7}$ (i.e., once a week).
	
	\textit{c.} Each group in the model is assigned to one of three geographical clusters. In the social interaction environment, agents in the same cluster interact with each other. Based on IDU demographic data from \cite{ambekar2008size}, we impose the condition that only agents between the ages of 18 and 32 years can be IDUs, and the maximum duration for which an agent engages in injecting drug use is 3 years. %From \cite{ambekar2008size} , we calculate the percentage of IDUs who are between 18 and 29 years using:
	% \begin{equation*}
	%     \frac{\sum_{district} N_{district} \times g_{district}}{\sum_{district} N_{district}} 
	% \end{equation*}
	% Here, $g_{district}$ is the percentage of IDUs in a given district who are between the ages of 18 and 29 years.
	% We also calculate the duration of one remaining and IDU using:
	% and 29 years using:
	% \begin{equation*}
	%     \frac{\sum_{district} N_{district} \times D_{district}}{\sum_{district} N_{district}} 
	% \end{equation*}
	% Here, $D_{district}$ is the duration for which a person remains an IDU in a given district. The data for parameter was also obtained from \cite{ambekar2008size}.
	
	\textit{d.} An interaction between a non-IDU and an IDU can lead to a non-IDU becoming an IDU with probability $p_{inf}$. As documented in \cite{ambekar2008size}, the proportion unemployed among persons aged 18-29 years differs significantly between IDUs and non-IDUs ($p_{ue}^{IDU}$ and $p_{ue}^{gen}$ in the equations below). Incorporating this factor may allow simulation-based design of interventions for IDUs that take their employment status into consideration. Therefore, as given in equation~\ref{eqid} below, we decided to calculate $p_{inf}$ as a weighted average of the probabilities of influence for employed ($p_{inf}^{e}$) and unemployed persons ($p_{inf}^{ue}$), respectively.
	
	\begin{equation}
		\label{eqid}
		p_{inf}^{ue} \times p_{ue}^{gen} \;+\; p_{inf}^{e} \times (1 - p_{ue}^{gen}) = p_{inf}
	\end{equation}
	
	As discussed in \cite{das2019agent}, we assume that $p_{inf}^{ue}$ and $p_{inf}^{e}$ are related to each other per the ratio of the proportions of unemployed persons among IDUs and the general population, respectively (expressed below in equation~\ref{eqidr}).%Here, $p_{inf}^{ue}$ and $p_{inf}^{e}$ are the probabilities that unemployed non-IDUs and employed non-IDUs in the age group 18-29 can be influenced into becoming IDUs, respectively.
	
	\begin{equation}
		\label{eqidr}
		\frac{p_{inf}^{ue}}{p_{inf}^{e}} = \frac{p_{ue}^{IDU}}{p_{ue}^{gen}}
	\end{equation}
	
	We could not find literature that reported estimates for $p_{inf}$, $ p_{inf}^{ue}$, and $p_{inf}^{e}$. Therefore, we considered $p_{inf}$ to be our second calibration parameter, and calculated $p_{inf}^{ue}$ and $ p_{inf}^{e}$ from equations~\ref{eqid} and~\ref{eqidr} once the value of $p_{inf}$ is estimated via the calibration process.
	
	\textit{e.} \cite{azim2008hiv} reported that IDUs interact in groups of 1-2.8 persons. We took the group size of interaction for IDUs to be 3 because we do not explicitly model HCV transmission through tattooing, which is a factor for HCV transmission in the region concerning our study \citep{chakravarti2013study}. Given that HCV transmission through tattooing is also via infected needles, similar to HCV transmission via injecting drug use, we assumed an IDU injecting group size of 3 (approximately the upper limit of the reported range) to implicitly include the effects of tattooing on HCV transmission.
	
	\textit{f.} \cite{ambekar2008size} also provided data regarding the proportion of IDUs that reported sharing needles at least once for the surveyed districts. We denote this as $s_{district}$ for a given district. Thus, we found the proportion of IDUs who reported sharing needles at least using the weighted average of the $s_{district}$ values for all surveyed districts, weighted using the population of each district. We obtained this value as 50.4\%. Thus, the probability of sharing needles during a given injecting drug use event is likely to be less than this value. This yielded a reasonable initial estimate for estimating the per-injecting-event needle sharing probability, which is the third calibration parameter. Note that if there is one infected agent in a network of IDUs engaging in needle sharing, then the per-event infection probability through injections determines whether an uninfected agent gets infected.
	
	\subsection{HCV Transmission in the Education Environment}
	This environment was included in the simulation to incorporate IDU-based interactions regardless of geographical considerations (interactions in the social interaction environment are assumed to occur only between agents in the same cluster), and to also facilitate future research on the effects of an awareness campaign conducted in educational environments on HCV epidemiology. This environment also incorporates the processes of conversion of non-IDUs to IDUs and of uninfected IDUs into infected IDUs. Although we remove geography-based restrictions on HCV transmission in this environment, the contribution of this environment to HCV prevalences is found to be very low as only 20\% of agents between the ages of 18 and 24 years avail of higher education in Punjab \citep{mha16}.
	
	% Table generated by Excel2LaTeX from sheet 'Fathoming algorithm 2'
	%\begin{table}[htbp]
	%  \centering
	% \caption{HCV transmission environment parameters.}
	% \begin{tabular}{|l|c|c|l|}
	% \hline
	% Parameter & Estimate & Relevant environment & Source \\
	% \hline
	% \hline
	% $N_{bt}$     & 0.023 & Medical environment & \cite{mhfw14}  \\
	% \hline
	% \hline
	% $N_{dp}$     & 0.982 & Medical environment &  \cite{stat18}\\
	% $p_{inj}$     & 0.018 & Medical environment &  \cite{cdc03}\\
	% \hline
	% $p_{bt}$     & 0.925 & Medical environment & \cite{patel2014estimating} \\
	%  \hline
	% $p_{sur}$     & 0.330  & Medical environment &  \cite{chant1994investigation}\\
	% \hline
	% $p_{ue}^{IDU}$     & 0.260  & Social interaction environment &  \cite{ambekar2008size}\\
	% \hline
	%  $p_{ue}^{gen}$     & 0.166 & Social interaction environment &  \cite{mha16}\\
	% \hline
	%\end{tabular}%
	%\label{paramtab}%
	%\end{table}%
	
	\section{MODEL CALIBRATION AND THE STOCHASTIC RULER METHOD}
	\label{modcal}
	
	We now describe the conceptualization of the calibration process as a simulation optimization problem. We conducted preliminary experimentation to determine the calibration parameter search space. To determine the search space for the per-event infection probability in the medical environment, we considered parameter estimates between the estimate of per-event infection probability through injections and per-event infection probability through surgeries. Similarly, adjustments were made to the per-event influence probability so that the IDU prevalence moves towards its calibration target. 
	
	We define the calibration problem in a general sense first, and then describe its application to our case. Let $m$ be the number of calibration variables and $n$ be the number of outcome variables. We define $x$, the $m$-tuple of calibration parameters as $x = (x_1,x_2, ..., x_m)$. The $n$-tuple of simulation outcomes to be calibrated to the calibration targets is defined as $y = (y_1,y_2, ..., y_n)$. Note that each $y_i = f_i(x), \; i = 1 \text{ to } n$, where $f_i$ represents the relationship between the $i^{th}$ simulation outcome and the calibration parameters $x$, implicitly given by the simulation. Correspondingly, we define the $n$-tuple of the calibration targets as $y^0 = (y_{1}^{0},y_{2}^{0}, ..., y_{n}^{0})$.
	
	A discrete simulation optimization problem typically takes the form: 
	\begin{equation}
		\label{optdef}
		\underset{x\; \in\; \mathbb{S}}{\min}\; E[g(x)]
	\end{equation}
	
	Here $g(x)$ is the output of a single replication of the simulation, $x$ are the decision variables, and $\mathbb{S}$ is the discrete solution space. In our case, we construct the following function that measures the sum of the absolute values of the distances between the simulation outcomes $y$ and the calibration targets $y^0$.   
	
	\begin{equation}
		\label{objdef}
		h(y) = \sum\limits_{i = 1}^{n} \left|1 - \frac{y_i}{y_{i}^{0}}\right|
	\end{equation}
	
	$h(y)$ is a function of the random variable $y$, and given that each of the $y_i = f_i(x) \; (i = 1 \text{ to }n)$, $h(y)$ in turn becomes a function of the calibration parameters $x$ - that is, $h(y) = h(y_1 = f_1(x), y_2 = f_2(x),\dots, y_n = f_n(x))$. A single replicate output of the simulation, which in equation~\ref{optdef} is represented by $g(x)$, in our case is given by the median of, say, $k$ values of $h(y)$. We denote the median of the $k$ values of $h(y)$ as $\hat{h}(y)$. We choose to define $g(x)$ in this manner (i.e., instead of setting $h(y)$ directly equal to $g(x)$) because of the high variance of the $y_i$. Further, we denote the simulation outcomes corresponding to $\hat{h}(y)$ as $\hat{y}$. Also, note that the $\hat{y}$ are functions of the calibration parameters $x$; that is, we can denote the simulation outcomes as $\hat{y}(x)$.
	
	For the HCV simulation model, we define $x_1$ to be the per-event infection probability in the medical environment, $x_2$ to be the per-event needle sharing probability, and $x_3$ to be the per-event influence probability. $y_1$ is the HCV antibody prevalence, $y_2$ is the HCV RNA prevalence and $y_3$ is the IDU prevalence. Thus, in our model, given that we estimate $y$ from the simulation outcomes corresponding to $\hat{h}(y)$ (i.e., $\hat{y}$), we can write $\hat{y}$ = ($\hat{y}_1$,$\hat{y}_2$,$\hat{y}_3$).%We can define the corresponding expected values as $E[Y_1]$, $E[Y_2]$, and $E[Y_3]$.
	
	%As the outputs of the simulation are stochastic, thus we are interested in the expected values of $Y$, and $h(Y)$. We define:
	%\begin{equation*}
	%    \hat{Y} = E(Y) \hspace{0.25cm} \& \hspace{0.25cm}\hat{h}(Y) = E(h(Y))
	%\end{equation*}
	
	For the HCV transmission simulation, we observe that as the per-event infection probability in the medical environment ($x_1$) and the per-event needle sharing probability ($x_2$) increase (decrease), the estimated values of the HCV antibody prevalence ($\hat{y}_1$) and HCV RNA prevalence ($\hat{y}_2$) also increase (decrease), but do not impact the expected value of the IDU prevalence ($\hat{y}_3$). As the influence probability ($x_3$) increases (decreases), $\hat{y}_3$ increases (decreases). Thus, we have a non-decreasing (monotonic) relationship between the decision variables $x$ and the estimated simulation outcomes $\hat{y}$. %This monotonic relationship is expected to be maintained between $X$ and the $\hat{Y}$ as well. 
	
	This is seen in the results of our preliminary experimentation (Table~\ref{tabprel} below) to determine the range of possible values for the calibration parameters as well. Table~\ref{tabprel} contains the lower and higher bounds on the calibration parameters that we determined through these preliminary experiments.
	
	% Table generated by Excel2LaTeX from sheet 'Fathoming algorithm 2'
	\begin{table}[htbp]
		\centering
		\caption{Results of preliminary experimentation.}
		\begin{tabular}{|c|c|}
			\hline
			$x = (x_1, x_2, x_3)$ & $\hat{y} = (\hat{y_1}, \hat{y_2}, \hat{y_3})$ \\
			\hline
			(0.035, 0.2, 1.9 $\times 10^{-5}$) & (1.17\%, 0.934\%, 0.087\%) \\
			\hline
			(0.037, 0.4, 2.3 $\times 10^{-5}$) & (5.01\%, 4.0\%, 0.13\%) \\
			\hline
		\end{tabular}%
		\label{tabprel}%
	\end{table}%
	
	Note that the values of $\hat{y}^0$ (estimates of calibration targets) for our model are (3.6\%, 2.6\%, 0.1\%), as obtained from \cite{sood2018burden}. If we define $x$ = $(0.035, 0.2, 1.9 \times 10^{-5})$ as $x^{l}$ and $x$ = $(0.037, 0.4, 2.3 \times 10^{-5})$ as $x^{r}$, and the corresponding $\hat{y}$ values as $\hat{y}^{l}$ and $\hat{y}^{r}$, then the `optimal' values of $x$, given by $x^* = (x_1^*, x_2^*, x_3^*)$ must satisfy the following relationship: 
	
	\begin{equation*}
		x^{l}_{1} \leq x^{*}_{1} \leq x^{r}_{1} \hspace{0.25cm} \& \hspace{0.25cm}  x^{l}_{2} \leq x^{*}_{2} \leq x^{r}_{2} \hspace{0.25cm} \& \hspace{0.25cm}  x^{l}_{3} \leq x^{*}_{3} \leq x^{r}_{3} 
	\end{equation*}
	
	Now, for each calibration parameter, we define:
	\begin{equation*}
		S(x_1) = \{x_{1}^{1}, x_{1}^{2}, ..., x_{1}^{k1}\} \hspace{0.25cm} \& \hspace{0.25cm}  S(x_2) = \{x_{2}^{1}, x_{2}^{2}, ..., x_{2}^{k2}\} \hspace{0.25cm} \& \hspace{0.25cm} ... \hspace{0.25cm}  S(x_m) = \{x_{m}^{1}, x_{m}^{2}, ..., x_{m}^{km}\} 
	\end{equation*}
	Here, $S(x_i)$ represents the set of possible values the $i^{th} \;(i = 1 \text{ to } m)$ calibration parameters can take, with $ki$ representing the cardinality of $S(x_i), \;i = 1 \text{ to } m$. The solution space $\mathbb{S}$ is formed by the Cartesian product of the $S(x_i)$. 
	%Let us assume without loss of generality that:
	%\begin{equation*}
	%    X_{i}^{1} < X_{i}^{2} < ... < X_{i}^{ki}, i = 1, 2, ..., m
	%\end{equation*}
	
	For our model, we define:
	\begin{equation*}
		\begin{aligned}
			&S(x_1) = \{0.035, 0.03525, 0.0355, 0.03575, 0.036, 0.03625, 0.0365, 0.03675, 0.037\}\\
			&S(x_2) = \{0.2, 0.25, 0.3, 0.35, 0.4\}\\
			&S(x_3) = \{1.9 \times 10^{-5}, 2.0 \times 10^{-5}, 2.1 \times 10^{-5}, 2.2\times 10^{-5}, 2.3 \times 10^{-5}\}\\
		\end{aligned}
	\end{equation*}
	%The set of all systems, A is obtained by using using the Cartesian product of $S(X_1)$, $S(X_2)$, ..., $S(X_m)$, i.e.,
	%\begin{equation}
	%    A = S(X_1) \times S(X_2) \times \hspace{0.25cm} ... \hspace{0.25cm}S(X_m)
	%\end{equation}
	Note that the elements in each of the above $S(x_i)$ are arranged in increasing order. Without loss of generality, we shall assume henceforth that $\forall$ $x_i^j, x_i^l \in S(x_i) \; (j \neq l \text{ and } j, l \in \{1,2,\dots, ki\},\; i = 1 \text{ to } m)$, $x_i^j < x^l_i \text{ if } j < l$.  Note also that $k1$ = 9, $k2$ = 5 and $km$ = $k3$ = 5, implying $\mathbb{S}$ is a set with cardinality 225.
	
	We now describe the application of the stochastic ruler method to calibrate the HCV transmission simulation. We shall not explain the stochastic ruler algorithm in detail due to space limitations, and we refer the author to \cite{yan1992stochastic} for a detailed description of the algorithm and its underlying definitions and assumptions. %Our explanation will be for the adaptation of the algorithm for the calibration of an agent-based simulation model and our model forms an example of this case.
	%\begin{itemize}
	The algorithm derives its name from the stochastic ruler $\theta$ against which a replicate output from the simulation is compared. We recall here that the output of a single replication of the simulation, $g(x)$, is set to be the median of, say, $k$ replicate values of $h(y)$ (defined in equation~\ref{objdef}), which we denote by $g(x) = \hat{h}(y)$. Note that the ideal (optimal) value of $E[\hat{h}(y)]$ is 0, implying that the expected value of the simulation outcomes are equal to the calibration targets in this case. Thus a value of $\hat{h}(y)$ equal to, for example, 0.3 implies an average percentage deviation of 10\% for a simulation outcome from its calibration target. Thus the lower limit $a$ of the stochastic ruler can be set as any value of $\hat{h}(y)$ below the value of $\hat{h}(y)$ corresponding to the maximum allowable average fractional deviation of the simulation outcomes from their calibration targets. We refer to this particular value of $\hat{h}(y)$ as the `threshold' of interest, and denote it by the symbol $\delta$. We experiment with four such thresholds - 0.45, 0.375, 0.3 and 0.2. We set $a$ to be 0.1, lower than the smallest threshold explored in this analysis.
	
	For determining the upper limit $b$ of the stochastic ruler, we run the simulation at the two extreme $x$ values $x^{l}$ and $x^{r}$, and set $b$ equal to the maximum of the $\hat{h}$ values corresponding to $x^{l}$ and $x^{r}$. In other words, $b = \max{(\hat{h}(y^{l}), \hat{h}(y^{r}))}$. We obtained 1.446 as the value of $\hat{h}(y^{l})$ and 1.229 as the value of $\hat{h}(y^{r})$, and therefore $b$ = 1.446.
	
	Next, we construct the neighborhood structure for each candidate solution $x$. We first define a neighbor set $N(x_i^j)$ for each $x_{i}^j \in S(x_i)$ $(j = 1 \text{ to } ki, \;i = 1 \text{ to } m$), in the following manner.
	
	\begin{equation*}
		\begin{aligned}
			& N(x_{i}^{j}) = \{x_{i}^{j-1}, x_{i}^{j}, x_{i}^{j+1}\},\hspace{0.25cm} 2 \leq j \leq (ki-1) \hspace{0.25cm} \\
			& N(x_{i}^{j}) = \{x_{i}^{ki}, x_{i}^{1}, x_{i}^{2}\},\hspace{0.25cm} j = 1 \hspace{0.25cm} \\
			& N(x_{i}^{j}) = \{x_{i}^{ki-1}, x_{i}^{ki}, x_{i}^{1}\}, \hspace{0.25cm} j = ki \\
		\end{aligned}
	\end{equation*}
	
	Then the collection of solutions forming the neighborhood of any solution $x = (x_1^j, x_2^j,\dots, x_m^j)$ is given by the Cartesian product of the $N(x_i^j)$ excluding $x$. In other words, $N(x) = \prod\limits_{i=1}^{m} N(x_{i}^{j}) - x$.
	
	We take the example of per-event infection probability in the medical facility, $x_1$, to illustrate the construction of the above neighborhood structure. For $x_1^3$ = 0.0355, $N(x_1^3)$ = $\{0.03525, 0.0355, 0.03575\}$; for $x_1^1$ = 0.035, $N(x_1^1)$ = $\{0.037, 0.035, 0.03525\}$; and for $x_1^{k1}$ = $x_1^9$ = 0.037, $N(x_1^9)$ = $\{0.03675, 0.037, 0.035\}$.
	
	Therefore, for any $x$, we have 26 neighbors. For example, if $x = (0.0355, 0.3, 2.0 \times 10^{-9}$), then $N(x_1)$ = $\{0.03525, 0.0355, 0.03575\}$, $N(x_2)$ = $\{0.25, 0.3, 0.35\}$, $N(x_3)$ = $\{1.9 \times 10^{-9}, 2.0 \times 10^{-9}, 2.1 \times 10^{-9}\}$, and $N(x) = N(x_1) \times N(x_2) \times N(x_3) - x$. %We eliminate the system $(0.0355, 0.3, 2.0 \times 10^{-9}$) from the set of neighbors as a system cannot be its own neighbor.
	
	We initiated the stochastic ruler method with $x^{l}$. The method involves selection of an appropriate candidate solution for the next iteration, where an iteration $t$ is defined as the random sampling of a neighbor (candidate solution) of the current solution and checking whether it can be set as the next estimate of the solution to the optimization problem. A candidate solution $z$, sampled from the neighborhood $N(x)$ of the current solution $x$ with probability $\frac{1}{|N(x)|}$, is selected as the next estimate of the optimal solution if all $M_t$ tests against samples from the stochastic ruler $\theta(a,b)$ `succeed'. For a neighbor to be selected as the system at the next iteration $t + 1$, it has to pass $M_t$ number of tests, where each test involves generating a replicate value of $g(z) = \hat{h}(y_z)$ and a sample $\theta \sim \theta (a,b)$, and then checking whether $\hat{h}(y_z) \leq \theta$. If $\hat{h}(y_z) \leq \theta$ (i.e., a `success'), then another test is conducted with new samples $\hat{h}(y_z)$ and $\theta$ until one of the tests is unsuccessful or all $M_t$ tests are successful. If $\hat{h}(y_z) > \theta$ (an unsuccessful test), then $t := t + 1$, and the solution at iteration $t$ is retained as the estimate of the solution at iteration $t+1$. Note that if all tests succeed, then $z \in N(x)$ is taken as the next estimate of the optimal solution, and $t:= t + 1$. $M_t$ must be selected such that it must be an non-decreasing function of $t$. We set $M_t$ equal to $\ceil{log(t+10)-log(5)}$.
	\begin{comment}
	\begin{equation*}
	M_t = \ceil{\frac{log(t+10)}{log(5)}} 
	\end{equation*}
	\end{comment}
	
	The method stops if $t > T$ or $\hat{h}(y) < \delta$, where $T$ is the computational budget set in terms of the number of iterations.
	%\end{itemize}
	
	The results of applying the stochastic ruler method are depicted in Table~\ref{tabsr} below. We used a computational budget of 40 iterations, given that generating a single realization of $\hat{h}(y)$ requires approximately 3 hours on a Intel $i7$ workstation with 3.3 GHz clock speed and 32 GB memory. Given this computational budget, we see that solutions that yield average deviations from the calibration targets that are less than the maximum allowable average fractional deviations of $10\%$ and $6.67\%$ (corresponding to $\delta$ values of 0.3 and 0.2) are not found. However, solutions that yield average percentage deviations less than threshold values of 15\% and 12.5\% are found in 15 iterations. Note that the same solution found by the SR method yields an $\hat{h}(y)$ that is less than both $\delta$ thresholds of 0.45 and 0.375 (i.e., it yields an average fractional deviation less than both 0.15 and 0.125 from the calibration targets), and hence the rows of Table~\ref{tabsr} appear mostly identical.
	
	% Table generated by Excel2LaTeX from sheet 'Final_results'
	\begin{table}[htbp]
		\centering
		\caption{Model calibration via the stochastic ruler method: results. Notes: $\delta_{avg}$ denotes the maximum allowable average fractional deviation, $\delta_{avg(o)}$ denotes the average fractional deviation obtained from the calibration process, and $t_f$ denotes the number of iterations to termination.}
		\begin{tabular}{|c|c|c|c|c|c|c|c|c|}
			\hline
			\multirow{4}[4]{*}{$\delta \; (\delta_{avg})$} & \multirow{4}[4]{*}{$\hat{h}(y)\; (\delta_{avg(o)})$} & \multicolumn{3}{c|}{     Prevalence outcomes } & \multicolumn{3}{c|}{Calibration parameters} & $t_f$ \\
			&       & \multicolumn{3}{c|}{(\% of population)} & \multicolumn{1}{c}{} & \multicolumn{1}{c}{} &       &  \\
			\cline{3-8}          &       & HCV   & HCV   & IDU   & $X_1$     & $X_2$     & $X_3$     &  \\
			&       & Antibody & RNA   &       &       &       &       &  \\
			\hline
			0.45 (0.15) & 0.360 (0.120) & 2.98  & 2.42  & 0.112 & 0.03525 & 0.25  & 2.3   & 15 \\
			\hline
			0.375 (0.125) & 0.360 (0.120) & 2.98  & 2.42  & 0.112 & 0.03525 & 0.25  & 2.3   & 15 \\
			\hline
		\end{tabular}%
		\label{tabsr}
	\end{table}%
	
	\section{SOLUTION SPACE TRUNCATION METHOD}
	
	In this section, we describe a method to reduce the size of the solution space - i.e., truncate it - by exploiting the monotonic relationship between the estimated simulation outcomes $\hat{y} = (\hat{y}_1, \hat{y}_2,\dots,\hat{y}_n)$ and the calibration parameters $x_j \; (j = 1 \text{ to } m)$. We refer to this approach as the solution space truncation ($SST$) approach. We note here that when used in the stochastic ruler method, $\hat{y}$ represents the outcome of a single `replication' of the simulation (even though it actually represents the simulation outcomes corresponding to the median of $k$ replicate values of $\hat{h}(y)$). For the application of the SST approach, while we generate $\hat{y}$ in the same manner as for the stochastic ruler method, we also we utilize the fact that $\hat{y}$ is the median of, say, $k$ replicate values of the $y$. This is seen in Assumption~\ref{monoas} below, where we formalize the monotonic relationship between the $\hat{y}$ and the $x_j \; (j = 1 \text{ to } m)$.
	
	\begin{assumption}
		\label{monoas}
		If $x^{1}_j$ $\leq$ $x^{2}_j$, then $\hat{y}^{1}$ $\leq$ $\hat{y}^{2}$, for $j = 1 \text{ to } m$. Here, $x^{1}_j$ and $x^{2}_j$ $\in S(x_j)$ $(j = 1 \text{ to } m)$ and $\hat{y}^{1}$ and $\hat{y}^{2}$ are simulation outcomes corresponding to $\hat{h}(y^1)$ and $\hat{h}(y^2)$.
	\end{assumption}
	%Each variant involves evaluation of a desired output function for certain values of $X$. The evaluations are c till a certain condition defined on the desired output function is satisfied by the $x$ selected. We explain how to select the values of $X$ and develop the condition on the desired output functions later in this section.
	
	We begin the SST approach with $x^{l}$, and then systematically move to other solutions $x^t$, where $x^t \in \mathbb{S}$ and $t \in \{1,2,..,|\mathbb{S}|\}$. We remind readers here that: (a) $|\mathbb{S}| = \prod\limits_{j = 1}^m |S(x_j)|$, and (b) we assume that the elements $x^1_j, x^2_j,\dots, x^{kj}_j$ of each $S(x_j)$ are indexed in ascending order of magnitude; that is, if $l < k$ then $x^l_j < x^k_j$ for $l \neq k \text{ and } j = 1 \text{ to } m$. Further $kj = |S(x_j)|$.
	
	$x^{t+1}$ is obtained by incrementing the index $j$ of each component $x_j$ of the current $x^t$ by one. For example, if $x^t = (x^t_1,x^t_2,...,x^t_m)$, then $x^{t+1} = (x^{t+1}_1,x^{t+1}_2,...,x^{t+1}_m)$. For each solution $x^t \in \mathbb{S}$, we generate the corresponding $\hat{y}^t$, evaluate $\hat{y}^t$ using a criterion that we develop below, and move on to $x^{t+1}$ if $\hat{y}^t$ satisfies said criterion. Starting from $x^l$, if a given $x^t$ does not satisfy this criterion, we stop, and set $x^l = x^{t-1}$. We then start the second part of the SST approach by beginning with $x^t = x^r$ and decreasing the indices of the components of $x^t$ by one and evaluating the corresponding $\hat{y}^t$. We terminate the SST approach when $\hat{y^t}$ does not satisfy the evaluation criterion, and set $x^r = x^{t+1}$. This yields a redefinition of the solution space $\mathbb{S}$ with new `boundary' solutions $x^l$ and $x^r$. We describe the criterion for evaluating the $\hat{y}^t$ and the consequent process of redefining $\mathbb{S}$ below.
	
	As part of defining the $\hat{y}^t$ evaluation criterion, we define the optimal solution to the simulation optimization problem representing the calibration process as follows.
	
	\noindent \textbf{Definition.} $x^*$ is the optimal solution of problem~\ref{optdef} if $E[\hat{y}(x^*)]$ equals the calibration targets $y^0$.
	
	We also make the following assumption regarding the variance of $\hat{y}^t$ $\forall \; x^t \in \mathbb{S}$.
	
	\begin{assumption}
		\label{varas}
		The variance of the simulation outcome estimator $\hat{y}^t$ is small in comparison with the difference between $E[\hat{y}^t]$ and $E[\hat{y}^{t+1}]$ $\forall \; 1 \leq t \leq |\mathbb{S}| - 1$.
	\end{assumption}
	Assumption~\ref{varas} is required because we define the evaluation criterion only in terms of $\hat{y}^t$, and do not consider its variance. Determining the impact of the variance of $\hat{y}^t$ on the effectiveness of the SST approach is an important avenue of future research. However, we believe it is a reasonable assumption at this stage because for our case, the variance of $\hat{y}^t$ is small given the median-based definition of $\hat{y}^t$.
	
	Using Assumptions~\ref{monoas} and~\ref{varas} and the definition of $x^*$, we are now in a position to state the following proposition that forms the basis for the evaluation criterion used in the SST approach. 
	
	\begin{proposition}
		If $\hat{y}^t$ $\prec$ $y^{0}$ then $x^*$ $\not\in$ $\mathbb{B}_t$, where $\mathbb{B}_t$ = \{$x \in \mathbb{S}$ | $x \preccurlyeq x^t$\}, where $\prec$ and $\preccurlyeq$ indicate element-wise comparisons. 
		Similarly, if $\hat{y}^t \succ y^0$ then $x^*$ $\not\in$ $\mathbb{C}_t$, where $\mathbb{C}_t$ = \{$x \in \mathbb{S}$ | $x \succcurlyeq x^t$\}, where $\succ$ and $\succcurlyeq$ indicate element-wise comparisons. 
		
		%\textit{Similarly, if} $\hat{Y}^{current}$ $\succ$ $\hat{Y}^{0}$ \textit{then an optimal solution to the calibration problem defined by objective function} $(1)$ \textit{and solution space} $\mathbb{S}$ \textit{does not lie in the set} $\mathbb{C}$ \textit{defined as} $\mathbb{C}$ = \{$X$| $X$ \succcurlyeq $X^{current}$\}, \textit{where} $\succ$ \textit{and} $\succcurlyeq$ \textit{indicate element-wise greater than and greater than or equal to inequalities respectively. $X^{current}$ is the current system state for which we obtain the $\hat{Y}$} value as $\hat{Y}^{current}$.
	\end{proposition}
	\textbf{Proof.} Let $\hat{Y}^{t}$ $\prec$ $\hat{Y}^{0}$, where $\hat{Y}^{0}$ is the value of $\hat{Y}$ at optimal point $X^{*}$. Thus, due to the non-decreasing relationship between $\hat{Y}$ and $X$, $X^{t}$ $\prec$ $X^{*}$. Let $X^{B_t}$ $\in$ $B_{t}$. Then, by definition of $B_{t}$, $X^{B_t}$ $\prec$ $X^{t}$ $\prec$ $X^{*}$. Again, by the non-decreasing relationship between $\hat{Y}$ and $X$, we have $\hat{Y}^{B_t}$ $\prec$ $\hat{Y}^{t}$ $\prec$ $\hat{Y}^{0}$. Then, by definition of $Y$, we have, $\hat{Y}^{B_t}_{1}$ $<$ $\hat{Y}^{t}_{1}$  $<$ $\hat{Y}^{0}_{1}$, $\hat{Y}^{B_t}_{2}$ $<$ $\hat{Y}^{t}_{2}$ $<$ $\hat{Y}^{0}_{2}$, ..., $\hat{Y}^{B_t}_{n}$ $<$ $\hat{Y}^{t}_{n}$ $<$ $\hat{Y}^{0}_{n}$. Then, by the definition of $h(Y)$, we get $h(Y^{B_t})$ $>$ $h(Y^{t})$ $>$ $h(Y^{0})$ $=$ $0$. Thus, $X^{B_t}$ contains solutions worse than $X^{t}$, and hence is eliminated. Similarly, we will have $X^{C_t}$ $\succ$ $X^{t}$ $\succ$ $X^{*}$ and $\hat{Y}^{C_t}$ $\succ$ $\hat{Y}^{t}$ $\succ$ $\hat{Y}^{0}$, where $X^{C_t}$ $\in$ $C_t$. This leads to $\hat{Y}^{C_t}_{1}$ $>$ $\hat{Y}^{t}_{1}$ $>$ $\hat{Y}^{0}_{1}$, $\hat{Y}^{C_t}_{2}$ $>$ $\hat{Y}^{t}_{2}$ $>$ $\hat{Y}^{0}_{2}$, ..., $\hat{Y}^{C_t}_{n}$ $>$ $\hat{Y}^{t}_{n}$ $>$ $\hat{Y}^{0}_{n}$. Again, by the definition of $h(Y)$, we have $h(Y^{C_t})$ $>$ $h(Y^{t})$ $>$ $h(Y^{0})$ $=$ $0$. This leads to elimination of solutions in $C_t$, completing the proof.
	
	In the first pass of the SST approach, we begin from $x^l$, and generate the $x^t$ by simultaneously incrementing the indices of $x^l$ by one. For each $x^t$, we generate the corresponding $\hat{y}^t$ and test whether $\hat{y}^t \prec y^0$. If the condition is satisfied, then we eliminate solutions belonging to the set $\mathbb{B}_t$. This is because, by Assumption~\ref{monoas}, we cannot obtain any value of $\hat{y}$ that is closer to $y^0$ than $\hat{y}^{t}$ using solutions in set $\mathbb{B}_t$. If $\hat{y}^t \succcurlyeq y^0$ and $\hat{y}^{t} \neq y^{0}$, then we move to $x^{r}$ and begin generating $x^t$ by simultaneously decrementing the indices of the components of $x^r$ by one. For each $x^t$, we generate the corresponding $\hat{y}^t$ and check whether $\hat{y}^t \succ y^0$. If the condition is satisfied, then we eliminate solutions belonging to the set $\mathbb{C}_t$ following a similar logic to the removal of solutions in the set $\mathbb{B}_t$ in the first pass of the SST approach. If $\hat{y}^t \preccurlyeq y^0$ and $\hat{y}^{t} \neq y^0$, then we terminate the SST approach. 
	
	Note that we also terminate the SST approach (first or second passes) if we exhaust the index sets of any one of the components of the $x^t$; that is, we terminate the pass if at any point $t > \min (k1, k2, ..., km)$ for that pass. %This is because for a certain component of $X$, we cannot make any further increase (decrease) in its value for the next evaluation in the rightward (leftward) movement on systems if we have made $min\{k1, k2, ..., km\}$ evaluations for that particular movement.
	The SST approach is summarized in Algorithms 1 and 2. In order to implement the SST approach computationally, we construct a matrix $A$ comprising all $x_t \in \mathbb{S}$. Each row of $A$ consists of a solution $x^t \in \mathbb{S}$, where $x^t \in \mathbb{R}^m$, and thus $A$ is an $|\mathbb{S} \times m|$ matrix. In our implementation, the first column represented the per-event infection probability in the medical environment, the second represented the per-event needle sharing probability, and third represented the per-event influence probability. 
	
	\begin{algorithm}
		\caption{Solution space truncation approach: first pass.}
		\label{alg1}
		\begin{algorithmic}
			\STATE Initialize with $A$, $y^0$
			\STATE Initialize with $x^t = x^l$
			\FOR{$t = 1 \text{ to } \min (k1,k2, ..., km)$}
			\STATE Set $x$ = $x^t$, that is, $x_{1} = x_{1}^{t}$, $x_{2} = x_{2}^{t}$, ..., $x_{m} = x_{m}^{t}$
			\STATE Generate $\hat{y}(x)$% or $\hat{H}(Y^{current})$
			\IF{$\hat{y}(x) \prec y^0$}
			\FOR{$i=1$ to $|\mathbb{S}|$}
			\IF{$A(i,1) \leq x_1 \land A(i,2) \leq x_2 \land...\land A(i,m) \leq x_m $}
			\STATE Remove row $i$ from $A$
			\ENDIF
			\ENDFOR
			\ELSE \STATE BREAK  
			\ENDIF
			\STATE $x^t = x^{t+1}$; that is: $x^t_1 = x^{t+1}_1$, $x^t_1 = x^{t+1}_2$,..., $x^t_m = x^{t+1}_m$
			\ENDFOR
		\end{algorithmic}
	\end{algorithm}
	
	\begin{algorithm}
		\caption{Solution space truncation approach: second pass.}
		\label{alg1}
		\begin{algorithmic}
			\STATE Algorithm 2. Solution space truncation approach: second pass.
			\STATE Initialize with $A_l$ ($A$ generated as output of the first pass), $y^0$
			\STATE Initialize with $s_{new}$ = number of rows of $A_l$
			\STATE Initialize with $x^t = x^r$
			\FOR{$t = 1 \text{ to } \min (k1,k2, ..., km)$}
			\STATE Set $x$ = $x^t$, that is, $x_{1} = x_{1}^{t}$, $x_{2} = x_{2}^{t}$, ..., $x_{m} = x_{m}^{t}$
			\STATE Generate $\hat{y}(x)$% or $\hat{H}(Y^{current})$
			\IF{$\hat{y}(x) \succ y^0$}
			\FOR{$i=1$ to $s_{new}$}
			\IF{$A_l(i,1) \geq x_1 \land A_l(i,2) \geq x_2 \land...\land A_l(i,m) \geq x_m $}
			\STATE Remove row $i$ from $A_l$
			\ENDIF
			\ENDFOR
			\ELSE \STATE BREAK  
			\ENDIF
			\STATE $x^t = x^{t-1}$; that is: $x^t_1 = x^{t-1}_1$, $x^t_1 = x^{t-1}_2$,..., $x^t_m = x^{t-1}_m$
			\ENDFOR
			\STATE Algorithm output: reduced solution space $A_{out} = A_l$
		\end{algorithmic}
	\end{algorithm}
	
	The output of applying Algorithms 1 and 2 on $\mathbb{S}$ (represented by the matrix $A$ in the above algorithms) yields the truncated solution space $A_{out}$. Applying the stochastic ruler on this truncated solution space yields improved results, as seen in Table~\ref{tabsst}. 
	
	% Table generated by Excel2LaTeX from sheet 'Final_results'
	\begin{table}[htbp]
		\centering
		\caption{Model calibration via the stochastic ruler method and the $SST$ approach: results. Notes: $\delta_{avg}$ denotes the maximum allowable average fractional deviation, $\delta_{avg(o)}$ denotes the average fractional deviation obtained from the calibration process, and $t_f$ denotes the number of iterations to termination.}
		\begin{tabular}{|c|c|c|c|c|c|c|c|c|}
			\hline
			\multirow{4}[4]{*}{$\delta \; (\delta_{avg})$} & \multirow{4}[4]{*}{$\hat{h}(y)\; (\delta_{avg(o)})$} & \multicolumn{3}{c|}{     Prevalence outcomes } & \multicolumn{3}{c|}{Calibration parameters} & $t_f$ \\
			&       & \multicolumn{3}{c|}{(\% of population)} & \multicolumn{1}{c}{} & \multicolumn{1}{c}{} &       &  \\
			\cline{3-8}          &       & HCV   & HCV   & IDU   & $X_1$     & $X_2$     & $X_3$     &  \\
			&       & Antibody & RNA   &       &       &       &       &  \\
			\hline
			0.45 (0.15) & 0.444 (0.148) & 2.74  & 2.23  & 0.094 & 0.03625 & 0.3   & 2.1   & 1 \\
			\hline
			0.375 (0.125) & 0.2788 (0.092) & 3.37  & 2.75  & 0.116 & 0.03625 & 0.35  & 2.2   & 6 \\
			\hline
			0.3 (0.1) & 0.2788 (0.092) & 3.37  & 2.75  & 0.116 & 0.03625 & 0.35  & 2.2   & 6 \\
			\hline
			0.2 (0.067) & 0.1464 (0.05) & 3.34  & 2.7   & 0.104 & 0.03575 & 0.35  & 1.9   & 29 \\
			\hline
		\end{tabular}%
		\label{tabsst}%
	\end{table}%

	We see that when the stochastic ruler is applied on the solution space represented by $A_{out}$, solutions that yield average deviations from the calibration targets that are less than all desired threshold average fractional deviations - 15\%, 12.5\%, 10\%, and 6.67\% - are found. Further, we see that for all threshold average fractional deviations save $6.67\%$, the number of iterations until termination reduce substantially. However, we note here that a certain amount of computational effort must be expended in the application of the SST approach itself, and hence the tradeoff in achieving improved calibration accuracy at the cost of this computational effort must be evaluated prior to applying the SST approach. In our case, the stochastic ruler method on the original solution space required approximately 7 days and 6 hours until termination (exhaustion of the computational budget). In comparison, application of the $SST$ algorithm required 9 hours, and as can be seen from Table~\ref{tabsst}, finding solutions that yielded average fractional deviations less than 10\% from the calibration targets were found in less than half the number of iterations than when working with the original solution space.
	
	\section{CONCLUSIONS} 
	In this work, we present a discrete simulation optimization approach towards the estimation of key parameters of a complex agent-based simulation of HCV transmission via a calibration process. We apply the stochastic ruler method to solve the simulation optimization conceptualization of this model calibration process, and within a pre-specified computational budget, find solutions that achieve acceptable average deviations of the simulation outcomes from their calibration targets. However, upon application of a method that we develop to reduce the size of the solution space using the monotonic nature of the relationship between the simulation outcomes and the calibration parameters, we find improved solutions at lesser computational expense. 
	
	The HCV transmission agent-based model that we use here to demonstrate our approach towards model calibration has high variance in its key simulation outcomes. This necessitated the use of an aggregate measure of simulation outcomes in the application of the stochastic ruler method, which added to the significant computational overhead of the calibration process. However, we anticipate that this approach towards simulation model calibration can be used for other simulations (e.g., discrete-event, Monte Carlo, or other agent-based simulations) in different contexts, which may be subject to less variance in their outcomes. In such situations, the computational overhead of this approach may be substantially lower. Hence, an avenue of future work involves exploring the use of this calibration for other simulations in different application settings.

	\if0\blind{
		\section*{Acknowledgements}
		The authors acknowledge the generous support from Prof. Devendra Dubey, Ms. Rachna Joshi and Mr. Kamal Rana for the technical support provided by them in form of the workstations on which we ran our simulation experiments.} \fi
	
	\bibliographystyle{chicago}
	\spacingset{1}
	\bibliography{IISE-Trans}
	
	\section*{AUTHOR BIOGRAPHIES}
	
	\noindent {\bf SOHAM DAS} is a Ph.D. student in the Department of Mechanical Engineering at the Indian Institute of Technology Delhi. His email address is 2019mez8426@mech.iitd.ac.in. \\
	
	\noindent \textbf{NAVONIL MUSTAFEE} is a Professor at the University of Exeter Business School. His email address is n.mustafee@exeter.ac.uk.\\
	
	\noindent {\bf VARUN RAMAMOHAN} is an Assistant Professor in the Department of Mechanical Engineering at the Indian Institute of Technology Delhi. His email address is varunr@mech.iitd.ac.in. \\
	
\end{document}